\documentstyle[prl,aps]{revtex}
\def\aap{{Astronomy and Astrophys.}}	

\def\mnras{{MNRAS}}

\def\ga{\gtrsim}

\def\eg{{ e.g.,\ }}
\def\ie{{ i.e.,\ }}

\twocolumn
\begin{document}
\draft

\title{ Spectral universality of strong shocks accelerating charged 
particles} \author{M.A. Malkov} \address{Max-Planck Institut f\"ur 
Kernphysik, D-69029, Heidelberg, Germany} 
 
\maketitle
\begin{abstract}
As a rule, the shock compression controls the spectrum of diffusively 
accelerated particles. We argue that this is not so if the backreaction 
of these particles on the shock structure is significant.  
We present a self-similar solution in which the 
accelerated particles change the flow structure near the shock so strongly
that the total shock compression may become arbitrarily large.  Despite 
this, the energy spectrum behind the shock is close to $E^{-3/2}$
independently of anything at all.

\end{abstract}
The first order Fermi acceleration 
at shocks was always regarded as one of those universal phenomena that 
result in a scale-invariant behavior of physical quantities.  In 
particular, this acceleration mechanism has predicted a power-law 
momentum distribution \cite{kr:ea,dru} similar to those  observed in diverse 
astrophysical objects.  There exists, however, another important 
characteristics, the energy content of accelerated particles.  
Generally speaking,this 
quantity makes the expectations of the scale invariance not so well 
grounded.  Whenever the shock accelerates efficiently in the sense of 
its energy conversion into energetic particles (and this is precisely 
what one anticipates in the context of the problem of cosmic ray 
origin), a more complicated flow structure must form in place of this 
shock.  Besides the shock itself, usually called subshock, it contains 
a long precursor in which the inflowing gas is gradually decelerated 
by the pressure gradient of energetic particles, in other words of the 
cosmic rays (CRs) produced in the entire shock transition, Fig.  1.  
One interesting aspect of this gas-CR coupling is an increase 
of the total compression ratio which should harden the spectrum and 
make the CR acceleration potentially much more efficient.  At the same 
time, the most attractive feature of this acceleration mechanism, the 
scale-free form of the spectrum is likely to be lost since the shock 
develops now its internal structure and particles with different 
momenta ``see'' different shock compression due to the strong momentum 
dependence of the CR diffusivity.  It is also well known (see \eg Ref.  
\cite{dru}) that this broadening of the shock transition softens the
spectrum.  The combined effect of these two oppositely acting factors
has been resisting any regular analytic calculations for a long time
(see, however, \cite{bla80} for a perturbative approach).

Recently \cite{m97a}, we have reduced the problem of the gas-CR 
coupling to one nonlinear integral equation.  Its solution does 
provide selfconsistently both the particle spectrum and the structure 
of the hydrodynamic flow.  At least in a certain phase space region 
this solution turns out to preserve the scale-invariance of the test 
particle results.  But the most surprising aspect of this solution is 
that in contrast to the well known linear solution for the particle 
momentum distribution \( f \propto p^{-q} \) where \( q=3r/(r-1) \) 
and \( r \) stands for the shock compression ratio, in the nonlinear 
regime \( q=3\frac{1}{2} \) independently of anything at all.  This 
means that when accelerated particles modify the shock strongly  
their spectrum and the flow profile mutually adjust in such a way that 
{\it all such shocks become universal accelerators producing 
essentially the same power-law spectra}.  They differ only in the 
cut-off momentum that is always limited by the size and/or the age of
the  shock.  The above result is, however, limited formally to
relativistic particles with the Bohm diffusivity \( \kappa \propto p \)
and relates to the solution (out of the three possible) with the
highest conversion efficiency of the flow energy to the CR energy.

\input boxedeps.tex     
\SetOzTeXEPSFSpecial
\SetDefaultEPSFScale{500}
\HideDisplacementBoxes 
\begin{figure}
\TrimRight{1.5cm}
\TrimLeft{1cm}
\TrimTop{1cm}
\TrimBottom{16.7cm}
 {\bBoxedEPSF{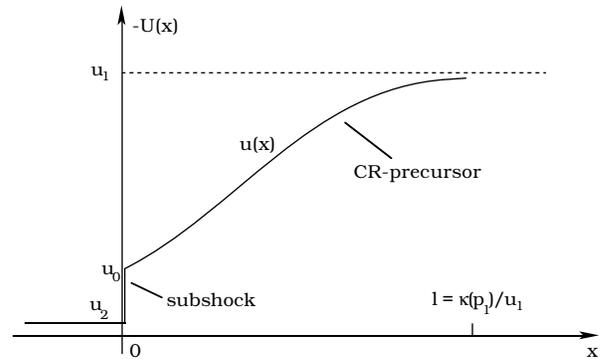}}	
	\caption{The flow structure in a strongly modified CR shock.}
	\protect\label{fig1}
\end{figure}
In this {\it letter}, using a simpler mathematical technique, we 
demonstrate that the spectrum universality persists also in the 
nonrelativistic region and for even more general form of diffusion 
coefficient, \eg \( \kappa \propto p^2/\sqrt{1+p^2/m^2 c^2} \).  The 
power-law index, being \( q=3\frac{1}{2} \) in the region \mbox{\( p 
\gg mc \)}, simply changes for \( q=4 \) in the region \( p \ll mc \) 
before it conforms to the local subshock compression \( r_{\rm s} \), 
i.e.  approaches \( q=3r_{\rm s}/(r_{\rm s}-1) \) at \( p \sim p_0 \ll 
mc \), where \( p_0 \) is the injection momentum \cite{m97a,mv95}.  
Remarkably, the \( q=3\frac{1}{2} \) index coincides with the plain 
test particle result for a strong shock in a purely thermal 
relativistic gas, whereas \( q = 4 \) coincides with that for a 
nonrelativistic gas.  All these in spite of the facts that the flow 
profile is strongly modified and the total compression ratio may be 
much higher than \( 4 \) (\( q=4 \)) and even \( 7 \) (\( 
q=3\frac{1}{2} \)) occurring in nonrelativistic and relativistic 
gases, respectively.  Moreover, the subshock compression ratio \( 
r_{\rm s} \) may be significantly lower than 4 and cannot be 
responsible for \( q=4 \) spectrum in the nonrelativistic momentum 
range.

The standard formulation of the problem includes the 
diffusion-convection equation constrained by conservation of the 
fluxes of mass and momentum (see \eg Ref.\ \cite{dru}).  We assume 
that a strong CR-modified shock propagates in the positive \( x 
\)-direction in a cold medium with the mass density \( \rho_1 \).  In 
the shock frame of reference the steady mass flow profile is defined 
as follows \( U(x)= -u(x), \, x \ge 0 \) and \( U(x)= -u_2, \, x < 0 
\), where \( u_2 \) is the (constant) downstream mass velocity, \( 
u(0+)=u_0 \), and \( u(x) \to u_1=const\), as \( x \to \infty \), Fig.  
1.  The equations read
\begin{eqnarray}
	\frac{\partial}{\partial x} \left( u g + \kappa(p) \frac{\partial 
g}{\partial x}\right) & = & \frac{1}{3} \frac{du}{dx} p\frac{\partial 
g}{\partial p}, 
	\protect\label{c:d} \\
	\rho u & = & \rho_1 u_1,
	\protect\label{c:e}  \\
	P_{\rm c}+\rho u^2  & = & \rho_1 u_1^2 , \quad x>0
	\protect\label{ber}
\end{eqnarray}
Here the number density of CRs is normalized to \( 4\pi gdp/p \).  The 
upper limit $p_1$ stands for a boundary in the momentum space 
(cut-off) beyond which particles are assumed to leave the system 
instantaneously ($g\equiv 0, \, p>p_1$).  In the downstream medium, \( 
x < 0 \), the only bounded solution is \( g=G(p) \equiv g(p,x=0) \).  
In Eqs.~(\ref{c:e},\ref{ber}) \( \rho(x) \) is the mass density, \( 
\rho_1 =\rho(\infty) \), \( P_{\rm c} \) is the CR pressure
\begin{equation}
	P_{\rm c}(x)= \frac{4\pi}{3} mc^2 \int_{p_0}^{p_1}\frac{p 
dp}{\sqrt{p^2+1}} g(p,x)
	\label{P_c}
\end{equation}
The particle momentum \( p \) is normalized to \( mc \).  As 
indicated, Eq.  (\ref{ber}) is written in the region $x>0$ where we 
have neglected the contribution of the adiabatically compressed cold 
gas (i.e., particles with \( 0 < p < p_0 \)) confining our 
consideration to sufficiently strong shocks with \( M^2 \equiv \rho_1 
u_1^2/\gamma P_{\rm g1} \gg (u_1/u_0)^{\gamma} \), where \( \gamma \) 
is the specific heat ratio of the plasma and $P_{\rm g1} $ is the gas 
kinetic pressure far upstream, \ie at $x=\infty$ (see \cite{m97a} for 
a detailed discussion of this approximation).  The subshock strength 
can be obtained from the familiar Rankine-Hugoniot condition for the 
gas
\begin{equation}
	r_{\rm s} \equiv \frac{u_0}{u_2}=\frac{\gamma+1}{\gamma-1+ 
	2(u_1/u_0)^{\gamma+1}M^{-2}}
	\label{c:r}
\end{equation}

Introducing the flow potential \( \phi \), such as \( u=d\phi/dx \) we 
seek the solution of Eq.  (\ref{c:d}) in the form
\begin{equation}
	g=g_0(p)\exp \left\{-\frac{1+\beta}{\kappa(p)}\phi\right\}, \quad 
	x>0
	\label{c:d:sol}
\end{equation}
where 
\begin{equation}
	\beta(p) \equiv -(1/3)d \ln g_0/d \ln p 
	\label{beta}
\end{equation} 
This is the key substitution in our analysis.  Considering $u$ as 
$u(\phi)$, substituting Eq.(\ref{c:d:sol}) in Eq.  (\ref{c:d}) and 
separating the variables we obtain
\begin{equation}
	\phi\frac{du}{d\phi}=\lambda u
	\label{fi:s} 
\end{equation}
\begin{equation}
	p \frac{d \beta}{d p}=(1+\beta) \left(\frac{d \ln 
\kappa}{d \ln p}-\frac{3}{\lambda} \beta\right)
	\label{bet:s1}
\end{equation}
where \( \lambda \) is a constant. Eq. (\ref{fi:s}) may be readily 
integrated and yields for the flow potential
\begin{equation}
	\phi(x)=\phi_0^{-\lambda /(1-\lambda)} \left[(1-\lambda) u_0 x 
+\phi_0\right]^{1/(1-\lambda) }
	\label{fi:s1}
\end{equation}
where \( \phi_0=\phi(0) \) is another constant (see below).  

The solution for the particle spectrum $g_{0}(p)$ to be obtained from 
the system (\ref{beta},\ref{bet:s1}) depends on two further 
integration constants.  These are the magnitude and the slope of $g_0$ 
at the injection momentum $p=p_0$, \ie $g_0(p_0)$ and $\beta_0\equiv 
\beta(p_0)$, respectively, provided by an ``injection'' solution 
\cite{mv95}.  Throughout this paper we shall consider $g_0(p_0)$ as 
given.  The constant $\beta_0$ may also be determined from merely the 
subshock strength (see also ref.  \cite{m97a} and below).  It is 
straightforward to verify that the following expression is the first 
integral of the system (\ref{beta},\ref{bet:s1})
\begin{equation}
	g_0(p)\left(\frac{\kappa}{1+\beta}\right)^\lambda 
	=g_0(p_0)\left(\frac{\kappa_0}{1+\beta_0}\right)^\lambda
	\label{first}
\end{equation}
where $\kappa_0\equiv\kappa(p_0)$.  The solution for $g_0$ is given by
\begin{eqnarray}	
	g_0(p) & = & g_0(p_0) \left(\frac{p}{p_0}\right)^3 \left[ 1+ 
	\right.  \nonumber \\
	 & & \left.  3 \frac{\beta_0+1}{\lambda\kappa_0}p_0^{-3/\lambda} 
	 \int_{p_0}^{p} \kappa(p') p'^{3/\lambda -1} d p' 
	 \right]^{-\lambda}
	\label{g0:sol}
\end{eqnarray}
For $p$ not too close to $p_0$, more precisely for 
$(\kappa/\kappa_0)(p/p_0)^{3/\lambda} \gg 1$, the particle spectrum 
$g_{0} $ acquires a form determined by merely that of $\kappa(p)$ and 
``forgets'' its behavior at $p \sim p_0$.  Note that the latter (and 
thus the parameter $\beta_0$) is prescribed by the subshock strength, 
according to the relation $d \ln G/d \ln p=3 /(r_{\rm s}-1)$ at 
$p=p_0$, which can be recast with the help of (\ref{first}) as
\begin{equation}
	\beta_0 -\phi_0 \beta_0\frac{1+\beta_0}{\kappa_0}= \frac{1}{r_{\rm 
	s}-1}
	\label{bet0}
\end{equation}
If $p$ is large in the above sense and $\kappa(p)$ is a power-law, 
then the spectrum $g_0(p)$ must also be, $g_0(p) \propto 
\kappa^{-\lambda}(p)$ and
\begin{equation}
	\beta \simeq (\lambda/3) 
	d \ln \kappa/d \ln p 
	\label{bet:ap}
\end{equation}
(the last formula may be also obtained  from Eq.(\ref{bet:s1}) assuming
$\beta \simeq const$).  As we shall see, the parameter $\lambda$
depends on the scaling of $\kappa$ as well, and the most surprising
consequence of this dependence is that the resulting slope of $g_0(p)$
is, in fact, independent of $\kappa(p)$.

What we obtained so far is a one parameter family of exact solutions 
to equation (\ref{c:d}) that require a rather special class of the 
flow profiles $u(\phi)$.  It is by no means guaranteed that any of 
these solutions satisfy the Bernoulli's integral (\ref{ber}) (the 
continuity condition (\ref{c:e}) can be easily satisfied since it does 
not depend on $g(p>p_0)$).  Indeed, the only unspecified quantity in 
solution (\ref{c:d:sol},\ref{fi:s1},\ref{g0:sol}) that can potentially 
be used to satisfy the functional relation (\ref{ber}), is the 
constant $\lambda$ which is, generally speaking, not enough.  
Fortunately, the dependence of $P_{\rm c}(\phi)$ in Eq.(\ref{ber}) can 
be made consistent with the dependence $u(\phi) \propto \phi^\lambda$, 
following from Eq.(\ref{fi:s}), in a fairly large part of shock 
transition, provided that the parameter $\lambda$ is chosen properly.  
To demonstrate this we substitute Eq.(\ref{c:d:sol}) into 
Eqs.(\ref{ber},\ref{P_c}).  Using Eq.(\ref{c:e}), the Bernoulli's 
integral (\ref{ber}) rewrites
\begin{equation}
	u(\phi)+\zeta \int_{s_1}^{s_0}\frac{d s}{\beta(s)} 
	s^{\lambda-1}e^{-\phi s} \frac{p^2(s)}{\sqrt{1+p^2(s)}} = u_1
	\label{ber1}
\end{equation}
where we have introduced a new variable $s$ in place of $p$
\begin{equation}
	s=\frac{1+\beta}{\kappa}
	\label{s}
\end{equation}
and the limits $s_{0,1}=s(p_{0,1})$.  We have also used the first 
integral (\ref{first}), $g_0 \propto s^\lambda$.  The injection rate 
$\zeta$ is defined here as
\begin{equation}
	\zeta=\frac{4\pi\lambda}{9}\frac{mc^{2}}{\rho_1 u_1} \left(\frac{\kappa_0}
	{1+\beta_0}\right)^\lambda g_0(p_0)
	\label{inj}
\end{equation}
and the function $p(s) $ in Eq.(\ref{ber1}) should be determined from 
Eq.(\ref{s}).  We now make a ``standard'' assumption: \( \kappa(p) = K 
p^2 (1+p^2)^{-1/2} \), i.e., the mean free path of a particle is 
proportional to its Larmor radius (here $K$ is a reference 
diffusivity).  Then, Eq.(\ref{ber1}) rewrites
\begin{equation}
	u(\phi)= u_1-\frac{\zeta}{K} 
	\int_{s_1}^{s_0}\left(1+\frac{1}{\beta(s)}\right) s^{\lambda-2} 
	e^{-\phi s} d s
	\label{ber2}
\end{equation}
According to Eqs.(\ref{bet:ap},\ref{s}), $\beta(s) $ is a very simple 
function, taking in the most part of its domain nearly constant and 
relatively close values, $\beta\simeq \lambda/3$ for $K s \ll 1$ and 
$\beta \simeq 2\lambda/3 $ for $K s \gg 1$.  It changes monotonically 
between these limiting values.  Therefore, the last integral allows a 
straightforward asymptotic analysis.  Our goal here is to examine 
whether Eq.(\ref{fi:s}), according to which $d u/d \phi \propto 
\phi^{\lambda-1}$ is compatible with Eq.(\ref{ber2}).  We assume that 
$0<\lambda <1$, which will be confirmed later.  Differentiating 
Eq.(\ref{ber2}) with respect to $\phi$ and considering first the 
region $1/s_0 \ll \phi \ll 1/s_1$ we may obviously replace the lower 
limit by zero and the upper one by infinity.  From Eq.(\ref{ber2}) we 
then obtain
\begin{eqnarray}
	\frac{d u}{d \phi} & \simeq & 
	\frac{\zeta}{K\phi^\lambda}\int_{0}^{\infty}\left[1+ 
	1/\beta(\tau/\phi)\right] \tau^{\lambda -1}e^{-\tau} d \tau 
	\nonumber \\
	 & = & \frac{\zeta\Gamma(\lambda)}{K\phi^\lambda}\left[1+ 
	 1/\beta(\bar{\tau}/\phi)\right]
	\label{ber:appr}
\end{eqnarray}
where $\Gamma$ is the gamma function and $\beta(\tau/\phi) $ is
replaced in the last integral by its mean value at $\bar{\tau}/\phi$
with $\bar{\tau} \sim 1$.  As we have seen already the function
$\beta(\bar{\tau}/\phi) $ in the last expression varies slowly and it
is close to $\lambda/3 $ for $\phi > K$ and to $ 2\lambda/3 $ for $\phi
< K$.  Therefore, the $\phi$ dependence of $d u/d \phi$ is determined
by the factor $\phi^{-\lambda}$ and is indeed consistent with
Eq.(\ref{fi:s}), \ie with $ u \propto \phi^\lambda$ for $\lambda=1/2$.
Eq.(\ref{ber:appr}) becomes invalid for $\phi \ga 1/s_1$, since the
lower limit in Eq.(\ref{ber2}) cannot be replaces by zero in this
case.  The function $d u/d \phi$ cuts off as (see Eq.(\ref{ber2}))
\begin{equation}
	\frac{d u}{d \phi} \simeq \frac{\zeta }{K\sqrt{s_1} 
	\phi}\left(1+\frac{1}{\beta(s_1)}\right) e^{-s_1 \phi}
	\label{ber:appr1}
\end{equation}
The last asymptotics obviously matches the formula (\ref{ber:appr}) at 
$\phi \sim 1/s_1$ but it is no longer consistent with  
Eq.(\ref{fi:s}).  At the same time, this  
is a periphery of the shock transition where 
$u(\phi)$ exponentially approaches its limit $u_1$ at \( x \sim l 
\equiv \kappa(p_1)/u_1 \), Fig.  1.  This particular form of the 
cut-off of $d u/d \phi$ is entirely due to our assumption about an 
abrupt cut-off of the particle spectrum at $p=p_1$ ($s=s_1$ in 
Eq.(\ref{ber2})).  Therefore, there is no major physical reason to 
refine our solution in this region unless a more realistic model of 
particle losses in the region $p\sim p_1$ is adopted.  According to 
Eq.(\ref{fi:s1}) the flow profile in the internal part of the 
precursor is simply linear  ($\lambda = 1/2$)
\begin{equation}
	u(x)=u_0+\frac{u_0^2}{2 \phi_0}x
	\label{u:lin}
\end{equation}

An asymptotic approach resulting in a flow profile $u(x)$ that is
uniformly valid (for all $x > 0$) and coincides with eq.(\ref{u:lin})
for $0 < x < l$ has been developed in Ref.  \cite{m97a}.  It produces a
full set of Rankine-Hugoniot relations, determining the total (\(
u_1/u_2 \)) and the subshock (\( u_0/u_2 \)) compression ratios along
with the constant $\phi_{0}$ in Eq.(\ref{fi:s1}) as functions of the
Mach number, the injection rate and the upper cut-off momentum \( p_1
\).  These quantities have been calculated in Ref.  \cite{m97a} with a
help of nonlinear integral equation derived from the system
(\ref{c:d}-\ref{ber}).  This equation has three different solutions in
a certain parameter range (see also \cite{m97b}) with dramatically
different acceleration efficiencies \cite{f1}.  However, its derivation
utilizes a relativistic form $\kappa \propto p$ for all $p$ and a
similar form of the  pressure integral.  These assumptions
oversimplify the particle spectrum at $p<1 $ and we have dropped them
here.  According to the form of the particle spectrum with the index
$q\equiv 3(\beta +1) =3.5$ at high momenta $p > 1$ and the linear
profile of the flow velocity (\ref{u:lin}) the solution obtained in the
present paper can be identified with that of the Ref.  \cite{m97a}
having the highest efficiency of CR production. Note that this solution
becomes unique for sufficiently large $\zeta$ and $p_1$.  
In addition, it provides parameters that have not been determined in 
the present
treatment.  First of all the relation between the flow deceleration in
the precursor and in the subshock was found to be $(u_1-u_0)/(u_0-u_2)
\propto \zeta p_1$ which, in combination with Eq.(\ref{c:r}) determines
the unknown velocities $u_0$ and $u_2$ given the far upstream velocity
$u_1$.  Also the constant $\phi_0$ may be related to the precursor
length $l=\kappa(p_1)/u_1$ by $u_0^2/2\phi_0 \simeq u_1/l$. 

As we already emphasized the scale-invariance does not exist 
everywhere in the phase space as in any other solution of this kind 
applied to a real physical situation.  However, the scale invariant 
behavior in the region \(u_0 < u < u_1 \), \( p_0 < p < p_1 \) may 
span decades in both momentum and coordinate space in those 
astrophysical situations where the dynamical range $p_1/p_0$ is large.  
E.g., for typical supernova shock conditions one may expect $p_1/p_0 
\sim 10^8$.  It is also interesting to note that the downstream 
particle spectrum given by Eq.(\ref{first}) with the approximate 
formula (\ref{bet:ap}) for $\beta$ and being expressed in terms of 
kinetic energy $E$ rather than momentum exhibits a particularly 
uniform behaviour throughout the entire energy range, relativistic and 
nonrelativistic.  In a standard normalization $F(E)d E$ this spectrum 
has the form
$$
F \propto E^{-3/2} \sqrt{\frac{(E+1)(7E^2+14E+8)}{(E+2)^3}}
$$
where $E$ is measured in $m c^2$, and the spectral index is close to 
$1.5$ for all but injection energies (if the subshock is substantially 
reduced, see text below Eq.(\ref{g0:sol})).  We emphasize that its 
value does not depend on any parameters involved in these 
calculations.  This results from the mutual adjustment of the flow 
profile \( u(x) \) and the particle distribution in energy and in this 
sense the form of the spectrum is completely universal.

One may ask then what does the power-law index depend on?  It should 
be primarily the momentum dependence of the CR diffusivity \( 
\kappa(p) \) that we have specified in our treatment above.  To 
examine this idea we rescale $\kappa$ as follows $\kappa ' = 
\kappa^\alpha$, where $\alpha > 1/2$ and the rescaling of all primed 
variables below is induced by the above transformation of $\kappa$.  
According to Eq.(\ref{bet:ap}) the spectral slope $\beta$ is now to be 
replaced by $\beta ' =(\alpha\lambda ' /3)d \ln \kappa /d \ln p$.  
Recalculating $d u /d \phi$ in Eq.(\ref{ber:appr}) with these rescaled 
spectrum and CR diffusivity $\kappa ' $ we obtain $d u /d \phi \propto 
\phi^{1/\alpha-1-\lambda ' }$.  Since the formula $u \propto 
\phi^{\lambda '} $ (Eq.(\ref{fi:s})) holds, we deduce that the 
rescaled $\lambda ' = 1/2\alpha$.  Therefore the spectral slope 
$\beta$ remains actually unchanged, $\beta ' = \beta$.

We conclude that the spectral universality is insensitive to the 
scaling of the CR diffusivity \( \kappa(p) \), in other words, to the 
spectrum of the underlying MHD turbulence.  On the other hand, the 
velocity profile does depend on \( \alpha \).  From the above analysis 
we obtain \( u \propto x^{1/(2\alpha -1)} \) which also explains the 
condition $\alpha > 1/2$ \cite{f2}.  
There is a small deviation from this scaling at 
the distances $x$ corresponding to the diffusion length of particles 
with $p \sim 1$ since $\beta$ depends on $\phi$ in 
Eq.(\ref{ber:appr}) at $\phi \sim K$.

Summarizing the last results, when \( \kappa(p) \) rescales, so does 
the flow profile \( u(x) \) but \( \beta \) remains invariant.  It is 
not difficult to understand why this is so.  As usual in the Fermi 
process the spectral slope of course depends on the flow compression.  
But, since the flow is modified, a particle with momentum \( p \), 
samples not the total compression but only a compression accessible to 
it.  This compression is determined by the relation \( \phi(x) \propto 
\kappa^{\alpha} \) (Eq.  (\ref{c:d:sol})).  As we have shown, \( 
u(\phi) \propto \phi^{1/2\alpha} \).  Therefore, the flow compression, 
as seen by this particle diffusively bound to the shock front, scales 
as \( u/u_2 \propto \phi^{1/2\alpha} \propto \sqrt{\kappa(p)} \).  As 
this is independent of \( \alpha \) the index \( \beta \) must also 
be.

This work was done within the Sonderforschungsbereich 328 of the 
Deutsche Forschungsgemeinschaft (DFG).



\begin{thebibliography}{}
\bibitem{f1}
The multiplicity of solutions for nonlinearly accelerating shocks is 
not new.  It has been studied in the framework of a hydrodynamic 
description (see \cite{dru} and the references therein).  However, 
there are very important differences between kinetic solutions 
obtained in \cite{m97a},\cite{m97b} and their hydrodynamic 
counterparts mostly because the latter are limited to the case in 
which the particle diffusivity $\kappa$ is independent of momentum 
whereas the approaches in \cite{m97a} and in the present paper not 
only allow for but also exploit this dependence.
\end{thebibliography}
\end{document}